\documentclass[a4,11pt]{reportN}
\usepackage[brazil]{babel}      % para texto em Português

\usepackage[utf8]{inputenc}   % para acentuação em Português

\usepackage{graphics}
\usepackage{subfigure}
\usepackage{graphicx}
\usepackage{epsfig}
\usepackage[centertags]{amsmath}
\usepackage{graphicx,indentfirst,amsmath,amsfonts,amssymb,amsthm,newlfont}
\usepackage{longtable}
\usepackage{cite}
\usepackage[usenames,dvipsnames,svgnames,table]{xcolor}

\usepackage{multicol}
\usepackage{verbatim}

\begin{document}

%********************************************************
\title{Incorporação de Incertezas Numéricas para Validação de Modelos Não Lineares}
%Este trabalho foi apresentado no evento X, na cidade Y no
%ano Z.}}

\author
    { \Large{Igor Carlini Silva}\thanks{igorufsj@yahoo.com.br} \\
 %{\small Grupo de Controle e Modelagem, UFSJ, São João Del Rei, MG}\\ 
  \Large{Gabriel Hugo Álvares Silva}\thanks{gabrielhugo0701@hotmail.com} \\
    % {\small Grupo de Controle e Modelagem, UFSJ, São João Del Rei, MG}\\
   \Large{Samir Angelo Milani Martins}\thanks {martins@ufsj.edu.br} \\
   \Large{Erivelton Geraldo Nepomuceno}\thanks {nepomuceno@ufsj.edu.br} \\
 {\small Grupo de Controle e Modelagem - GCOM, UFSJ, São João Del Rei, MG,\\
  Departamento de Engenharia Elétrica, UFSJ, São João Del Rei, MG} }

\criartitulo

%\markboth{\hfill Write the title of your work here, concisely if
%necessary \hfill Write the last authors' name here} {Proceeding Series of the Brazilian
%Society of Computational and Applied Mathematics \hfill}

%Os métodos de validação de modelos disponíveis na literatura não abrangem os erros computacionais.
\begin{abstract}
{\bf Resumo}.  O presente artigo tem por finalidade propor um método de validação de modelos que incorpore informações relativas aos erros no cômputo de funções computacionais durante a validação de modelos. Como estudos de caso, utilizam-se dois modelos identificados dos sistemas Mapa Seno e Circuito de Duffing-Ueda e os métodos de validação de modelos RMSE e MAPE. Pôde-se notar que, na 65ª iteração para o Mapa Seno, por exemplo, a diferença entre o método proposto e o da literatura, considerando o índice RMSE, é de 34\%.

%casos, o erro computacional apresenta influência, onde a diferença entre o método proposto e da literatura é de 0,4799 e $1,5099 \times 10^{-14}$, respectivamente.

\noindent
{\bf Palavras-chave}. Computação Aritmética, Validação de Modelos, Sistemas Caóticos, Identificação de Sistemas.
\end{abstract}

\section{Introdução}

Os sistemas dinâmicos experimentais são, a rigor, não lineares. Por esta razão, seu estudo e entendimento é de extrema importância. Para realizar a análise de tais sistemas, é necessária a obtenção de um modelo matemático que os represente. O modelo polinomial NARMAX (Nonlinear AutoRegressive Moving Average model with eXogenous input) vem sendo muito utilizado para representar sistemas dinâmicos não lineares \cite{billings2013nonlinear,chen1989}, sendo este um tipo de função recursiva na qual termos e parâmetros são selecionados cuidadosamente. Para a identificação de um modelo, podem ser utilizadas técnicas como: modelagem caixa-branca, caixa-cinza e caixa-preta \cite{Agu2007}.

Após a identificação de um modelo que representa um determinado sistema, é necessário verificar a sua confiabilidade, ou seja, o modelo deve ser validado. Entre as técnicas de validação de modelos disponíveis na literatura, pode-se citar os índices RMSE (Root Mean Square Error) e MAPE (Mean Absolute Percentual Error) \cite{Agu2007}. Além da identificação e modelagem de sistemas, experimentos numéricos são cruciais para o desenvolvimento de conhecimento sobre o comportamento de sistemas dinâmicos \cite{borah2017dynamics}, uma vez que, sendo descritos por funções recursivas, o resultado da iteração atual sempre depende do anterior \cite{rodriguesjunior2015}. A computação numérica usa aritmética de ponto flutuante na maior parte dos computadores sob o padrão IEEE 7542-2008. Como a computação numérica é uma aproximação dos números reais, utilizando um número finito de números, pequenos erros são gerados a cada iteração durante as simulações \cite{nepomuceno2014}.

No que diz respeito à propagação de erros, Galias \cite{galias2013} apresenta vários métodos computacionais para um estudo rigoroso de sistemas não lineares, considerando os erros por arredondamento, através de análise intervalar. Em \cite{hammel1987}, os autores mostram que, para parâmetros e condições iniciais específicos no mapa logístico, a distância entre duas órbitas próximas cresce geometricamente a cada iteração. Além disso, Nepomuceno e Martins \cite{nepomuceno2016lower} apresentam um método para avaliar o limite inferior do erro (Lower Bound Error – LBE) em funções recursivas, como ferramenta para aumentar a confiabilidade de simulações computacionais. Embora apresentem contribuições importantes no estudo da propagação de erros computacionais, não se tem conhecimento de trabalhos que avaliem a influência de tais erros em métodos de validação de modelos.

Portanto, este trabalho tem como finalidade verificar a influência dos erros computacionais na validação de modelos por meio de simulações computacionais, através da junção conceitual do LBE aos métodos existentes na literatura. A fim de verificar a eficácia da metodologia proposta, foram utilizados dois estudos de caso de modelos identificados dos sistema de Duffing-Ueda e Mapa Seno. O presente trabalho está organizado como segue. A segunda seção apresenta os conceitos preliminares relevantes para o desenvolvimento do trabalho. A terceira seção apresenta a metodologia adotada. Na quarta seção são expostos os resultados obtidos nos casos estudados e a quinta seção apresenta a conclusão do trabalho.    

\section{Conceitos Preliminares}

\subsection{Modelos NARMAX}

Os modelos polinomiais NARMAX são estruturas paramétricas capazes de representar o comportamento de uma ampla classe de sistemas dinâmicos não lineares reais. O modelo é matematicamente definido pela equação (1) \cite{chen1989}, onde $y_n$, $u_n$ e $e_n$, são, respectivamente, a saída, a entrada e o ruído no tempo discreto $n$ $\in$ $ \mathbb{N}$. Os parâmetros $k_y$, $k_u$ e $k_e$ são os máximos atrasos considerados e $F^\ell$ é uma função polinomial com estrutura a ser determinada de grau $\ell$.

\begin{eqnarray}
y_{n+1} = F^\ell[y_n,...,y_{n-1-k_y},u_n,...,u_{n-1-k_{u'}},  e_n,...,e_{n-1-k_e}]
\end{eqnarray}

\subsection{Índices RMSE (Root Mean Square Error) e MAPE (Mean Absolute Percentual Error)}

Dois índices para validação de modelos identificados frequentemente utilizados na literatura são os índices RMSE e MAPE \cite{Agu2007}. Estes índices podem ser expressos por:

\begin{minipage}{6cm}
\begin{eqnarray}
         \mathrm{RMSE}= \frac{\sqrt{\Sigma_{k=1}^N[y(k) - \hat y(k)]^2}}{\sqrt{\Sigma_{k=1}^N[y(k) - \bar y(k)]^2}} 
\end{eqnarray}
     
     \nonumber 

\end{minipage}\hfill
\begin{minipage}{6cm}
   \begin{eqnarray}
          \mathrm{MAPE}=\frac{1}{N}\sum_{k=1}^{N} {\left|\dfrac{\hat y(k)-y(k) }{y(k)}\right|}
   \end{eqnarray}
    \nonumber 

\end{minipage}

\vspace{0.5cm}

\noindent sendo $y(k)$ os dados mensurados do sistema, $\hat y$(k) a simulação do modelo e $\bar y$(k) a média dos dados de $y(k)$ para ambos os índices \cite{Agu2007}. Os índices apresentam restrições para serem computados, onde tanto o modelo quanto o sistema precisam ser compostos pelas mesmas entradas e condições iniciais. O índice RMSE representa a relação do somatório do erro quadrático entre o modelo e o sistema, e o somatório do erro quadrático entre o sistema e sua média \cite{Detec}. Já o índice MAPE representa a relação entre o módulo do somatório da diferença entre sistema e o modelo, e o produto do número de iterações (N) pelo sistema.

\subsection{Lower Bound Error}

O LBE é um método que tem por finalidade obter o limite inferior do erro computacional ocorrido a cada iteração de uma simulação. Este método foi desenvolvido por \cite{nepomuceno2016lower}, e sua representação matemática é dada por $2 \delta_{a,n}=|\hat x_{a,n} - \hat x_{b,n}|$.

Uma órbita é uma sequência de valores de um mapa, representada pelo subíndice $n$. Uma pseudo-órbita é uma aproximação de uma órbita. Neste caso, cada pseudo-órbita é representada por $\hat x$, cada uma com seus subíndices, $a$ e $b$, indicando suas extensões. Logo, $\{\hat x_{a,n}\}$ e $\{\hat x_{b,n}\}$ são duas pseudo-órbitas derivadas de duas extensões intervalares. Em uma simulação, se o limite inferior do erro for maior que a tolerância desejada, a metodologia utilizada deve ser reavaliada.

\section{Metodologia}

Uma vez que, em sistemas dinâmicos não lineares o estado futuro é extremamente dependente do estado atual, o erro propagado pode tomar dimensões consideráveis. A fim de considerar o erro numérico em métodos de validação de modelos, modificaram-se, dentre os métodos disponíveis na literatura, os índices RMSE e MAPE.

\subsection{LRMSE (Lower Root Mean Squared Error) e LMAPE (Lower Mean Absolute Percentual Error)}

Com o intuito de aperfeiçoar os métodos de validação existentes, levando em consideração os erros computacionais em suas equações, fez-se uso do conceito LBE para o desenvolvimento das equações matemáticas do novo método. Portanto, os novos índices RMSE e MAPE foram designados por $\mathrm{LRMSE}$ e $\mathrm{LMAPE}$, como mostrado pelas Equações (4) e (5). A obtenção destes novos índices foi possível devido ao fato de que os métodos de validação levam em consideração a simulação dos modelos matemáticos e têm por finalidade verificar o quão o modelo está próximo do sistema, além de apresentar os mínimos valores possíveis para os dados reais do sistema. Já o LBE mapeia o crescimento do limite inferior do erro. Portanto, o método proposto realiza a junção conceitual do LBE com os índices de validação, como o RMSE e o MAPE.

\begin{multicols}{2}
\tiny
\begin{equation}
  \mathrm{LRMSE}= \left\{\begin{aligned}
     \frac{\sqrt{\Sigma_{k=1}^N[y(k) - \hat y(k) \times(1-\delta_{a,n})]^2}}{\sqrt{\Sigma_{k=1}^N[y(k) - \bar y(k)]^2}}, \hspace{0.25cm} \hat y(k) < y(k)\\
     \frac{\sqrt{\Sigma_{k=1}^N[y(k) - \hat y(k) \times(1+\delta_{a,n})]^2}}{\sqrt{\Sigma_{k=1}^N[y(k) - \bar y(k)]^2}},  \hspace{0.25cm} \hat y(k) > y(k) 
  \end{aligned} \right.
\end{equation}

\begin{equation}
  \mathrm{LMAPE} =  \left\{ \begin{aligned}
     	\frac{1}{N}\sum_{k=1}^{N} \left|\dfrac{\hat y(k) \times(1-\delta_{a,n})-y(k) }{y(k)}\right|, \hspace{0.25cm} \hat y(k) < y(k)\\
      	\frac{1}{N}\sum_{k=1}^{N} \left|\dfrac{\hat y(k) \times(1+\delta_{a,n})-y(k) }{y(k)}\right|, \hspace{0.25cm} \hat y(k) > y(k)
\end{aligned} \right.
\end{equation}

\end{multicols}

A substituição de $\hat y(k)$ para $\hat y(k) \times (1 \pm \delta_{a,n})$ em ambos os índices tem por finalidade incluir as incertezas computacionais ocorridas a cada iteração por meio do índice LBE. Uma observação é realizada no que diz respeito a somar ou subtrair o termo $\hat y(k) \times \delta_{a,n}$ aos índices, cujo o objetivo é evitar os casos em que o erro possa aumentar e o índice diminuir, o que promoveria uma falsa interpretação. Com o intuito de verificar a eficiência do método proposto, foram utilizados dois estudos de caso de modelos identificados dos sistema de Duffing-Ueda e Mapa Seno. Para cada um deles, aplicou-se os métodos $\mathrm{LRMSE}$ e $\mathrm{LMAPE}$, de acordo com o procedimento descrito a seguir.

\begin{enumerate}
    \item Obtém-se os índices RMSE e MAPE através das Equações (2) e (3). Os termos $y(k)$ e $\hat y(k)$ representam os dados provenientes de $X_n$ e $G(X_n)$, sendo estes as equações do sistema e do modelo, respectivamente.
    \item Calcula-se os índices modificados
    \begin{enumerate}
        \item Primeiramente, calcula-se o LBE. As duas pseudo-órbitas consideradas foram $G(X_n)$ e $H(X_n)$, sendo estes o modelo e a extensão do sistema, respectivamente;
        \item Utilizando o LBE calculado no subitem anterior, calcula-se os índices $\mathrm{LRMSE}$ e $\mathrm{LMAPE}$, de acordo com as Equações (4) e (5). Os termos $y(k)$ e $\hat y(k)$ são os mesmos utilizados no item 1. 
    \end{enumerate}
    
    \item Um novo procedimento é empregado para obter os índices. Realizam-se os mesmos passos descritos anteriormente, mas substituindo $X_n$ por $G(X_n)$ e $G(X_n)$ por $H(X_n)$ nas equações. Utiliza-se o mesmo LBE calculado no item 2
(a). Esta etapa foi realizada para efeito de comparação e análise dos resultados entre os índices $\mathrm{LRMSE}$ e $\mathrm{LMAPE}$ obtidos.   
    
\end{enumerate}

\section{Resultados}
 
\subsection{Mapa Seno}

Através de manipulações nas partes sublinhadas das representações matemáticas dos sistemas, obtém-se duas extensões intervalares, sendo as Equações (6), (7) e (8) referentes ao Mapa Seno. As simulações ocorreram com as condições iniciais $X_p = 0,5$, sendo $p$ sua ordem, alterando de $p=0...3$. As comparações entre os resultados obtidos do RSME e $\mathrm{LRMSE}$, MAPE e $\mathrm{LMAPE}$ podem ser observadas na Figura 1.

\begin{multicols}{2}
\small
\noindent\begin{eqnarray}  
    X_n = 1,2\pi sen(x_{n-1})
\end{eqnarray}

\noindent\begin{eqnarray}
      G(X_n)& =& 2,6868X_n - \underline{0,2462X_n^3}\\
      H(X_n)& =& 2,6868X_n - \underline{(0,2462X_n)X_n^2}
\end{eqnarray}
\end{multicols}

\begin{figure}[ht!]
\subfigure[RMSE e $\mathrm{LRMSE}$ do Mapa Seno.]{
\includegraphics[width=7.25cm,height=5cm]{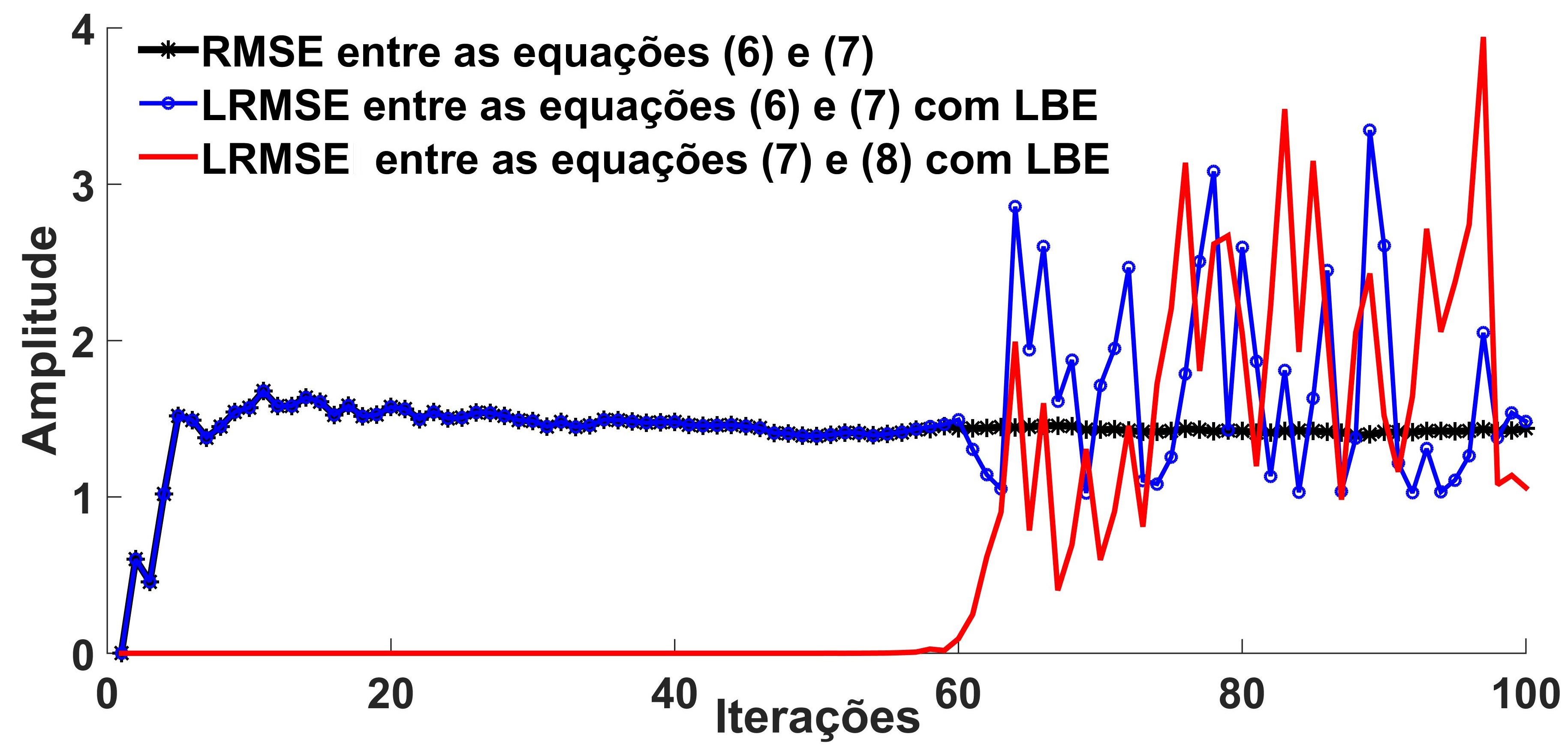}}
\subfigure[MAPE e $\mathrm{LMAPE}$ do Mapa Seno.]{
\includegraphics[width=7.25cm,height=5cm]{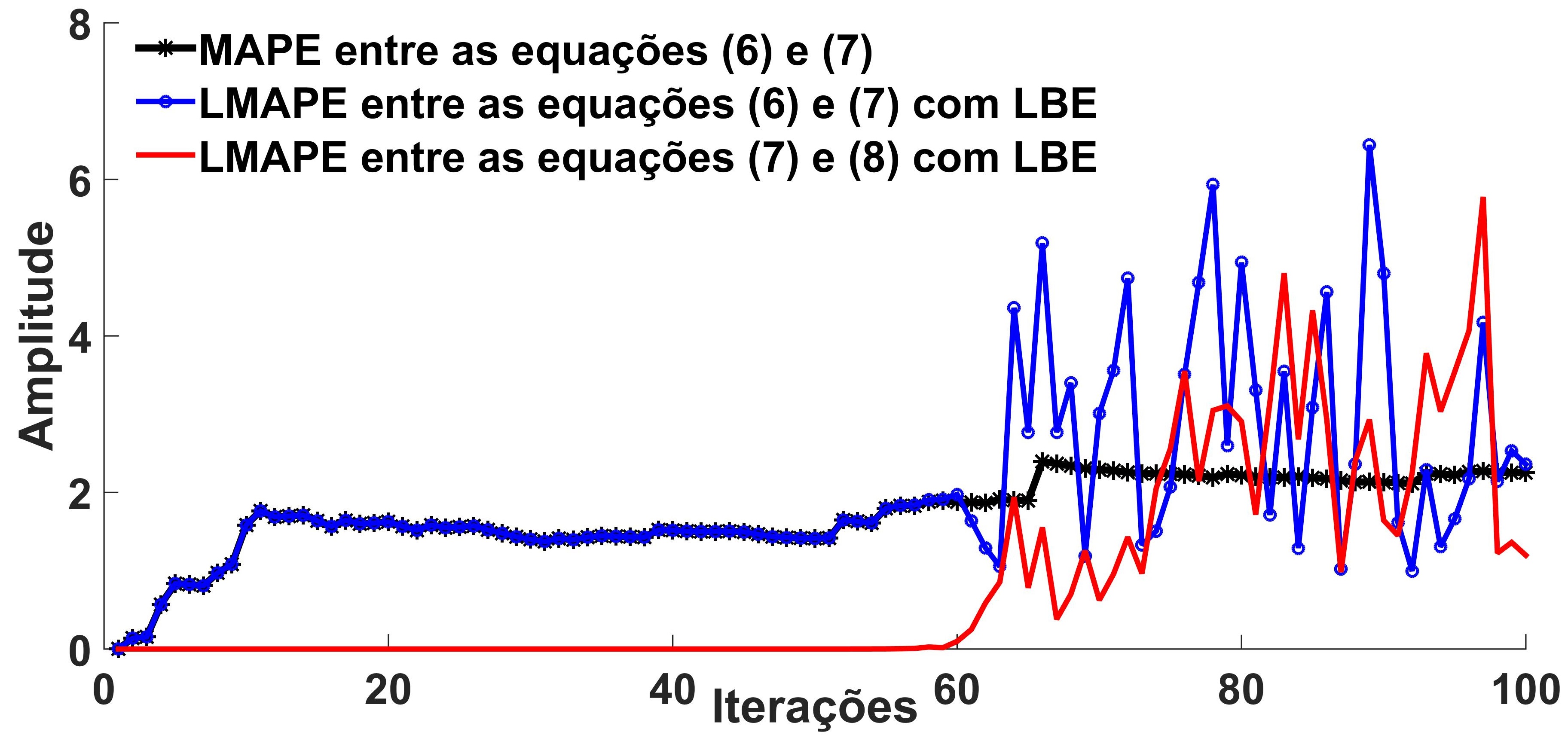}}
\caption{Análise dos resultados para os dois índices referente ao Mapa Seno.}
\end{figure}

Por meio da Figura (1), nota-se que o RMSE tende a permanecer em uma região. O $\mathrm{LRMSE}$, quando compara o sistema com o modelo, sendo as Equações (6) e (7), respectivamente, se inicia com o mesmo valor do RMSE mas, ao decorrer da simulação, o erro tende a crescer, sendo que na 65ª iteração apresenta uma diferença de 34\%, fazendo com que o índice oscile no entorno do RMSE. Na comparação entre o modelo e extensão, sendo as Equações (7) e (8), respectivamente, para o índice $\mathrm{LRMSE}$, ele inicia próximo de zero e posteriormente tende a oscilar no entorno do RMSE, apresentando uma diferença de 45\% na 65ª iteração. Isto se dá pelo fato da inclusão do erro aos métodos, uma vez que uma pequena pertubação causa resultados diferentes. Além de evidenciar que, apesar de obter um modelo que represente bem o sistema, ao realizar simulações no ambiente computacional, isto irá gerar resultados diferentes. A mesma análise pode ser feita para o índice $\mathrm{LMAPE}$, no qual a diferença é aproximadamente de 46\% e 58\%, respectivamente. 

%percebe-se que os valores se mantêm próximos de zero até um número significativo de iterações, o que não ocorre para o RMSE e MAPE originais. Isso ocorre pois as extensões matemáticas deveriam ser idênticas.

%Seno

% Sistema e modelo:  RMSEl = -0.494323053791910   -34.141709111838757%    MAPEl = -0.872290999148828  -46.061481248612914%
% Modelo e extensão: RMSEl =  0.662883483018838    45.783774150656605%   MAPEl =  1.111567581017614    58.696523682543621%

\subsection{Oscilador de Duffing-Ueda}
 
 A Equação (9) modela o sistema e \cite{Agu2007} apresenta a equação polinomial NARMAX do modelo. As pseudo-órbitas são obtidas através de modificações matemáticas ao modelo que estão sublinhadas, obtendo as Equações (10) e (11). Na simulação dos modelos NARMAX, quanto à extensão, adotou-se que as condições iniciais seriam $X_p=0$, sendo $p = 0...4$ e utilizou-se $U=Acos(nT_s)$, sendo $n \in N$, $T_s = \pi/60$, $A =10$, $k=1$ e $\mu= 0,25$. A Figura 2 apresenta a comparação entre os índices RMSE e $\mathrm{LRMSE}$, além da comparação entre MAPE e $\mathrm{LMAPE}$ .

\begin{figure}[ht!]
\subfigure[RMSE e $\mathrm{LRMSE}$ do Oscilador de Duffing-Ueda.\label{surreal1}]{
\includegraphics[width=7.25cm,height=3cm]{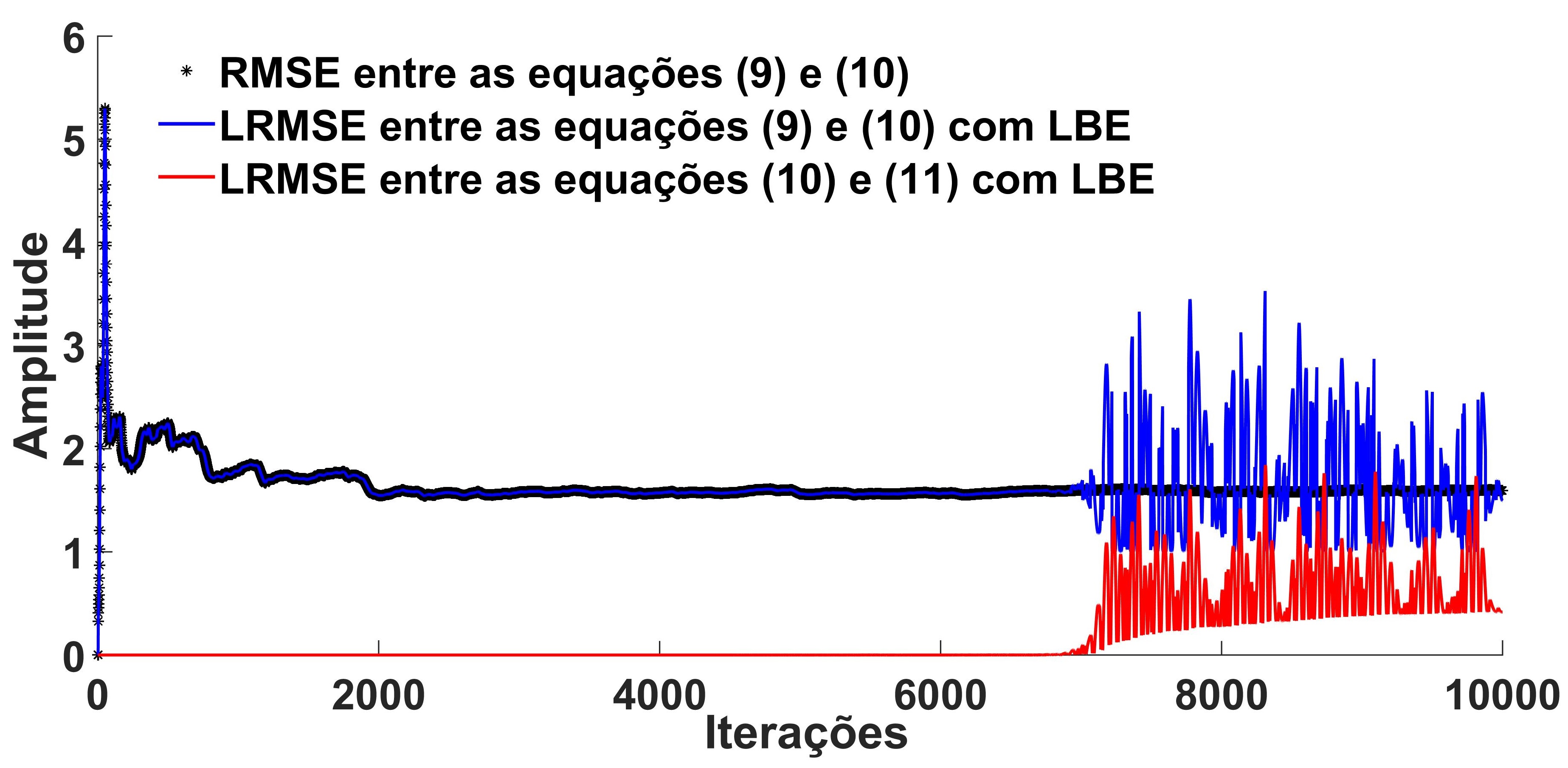}}
\subfigure[MAPE e $\mathrm{LMAPE}$ do Oscilador de Duffing-Ueda.\label{surreal2}]{
\includegraphics[width=7.25cm,height=3cm]{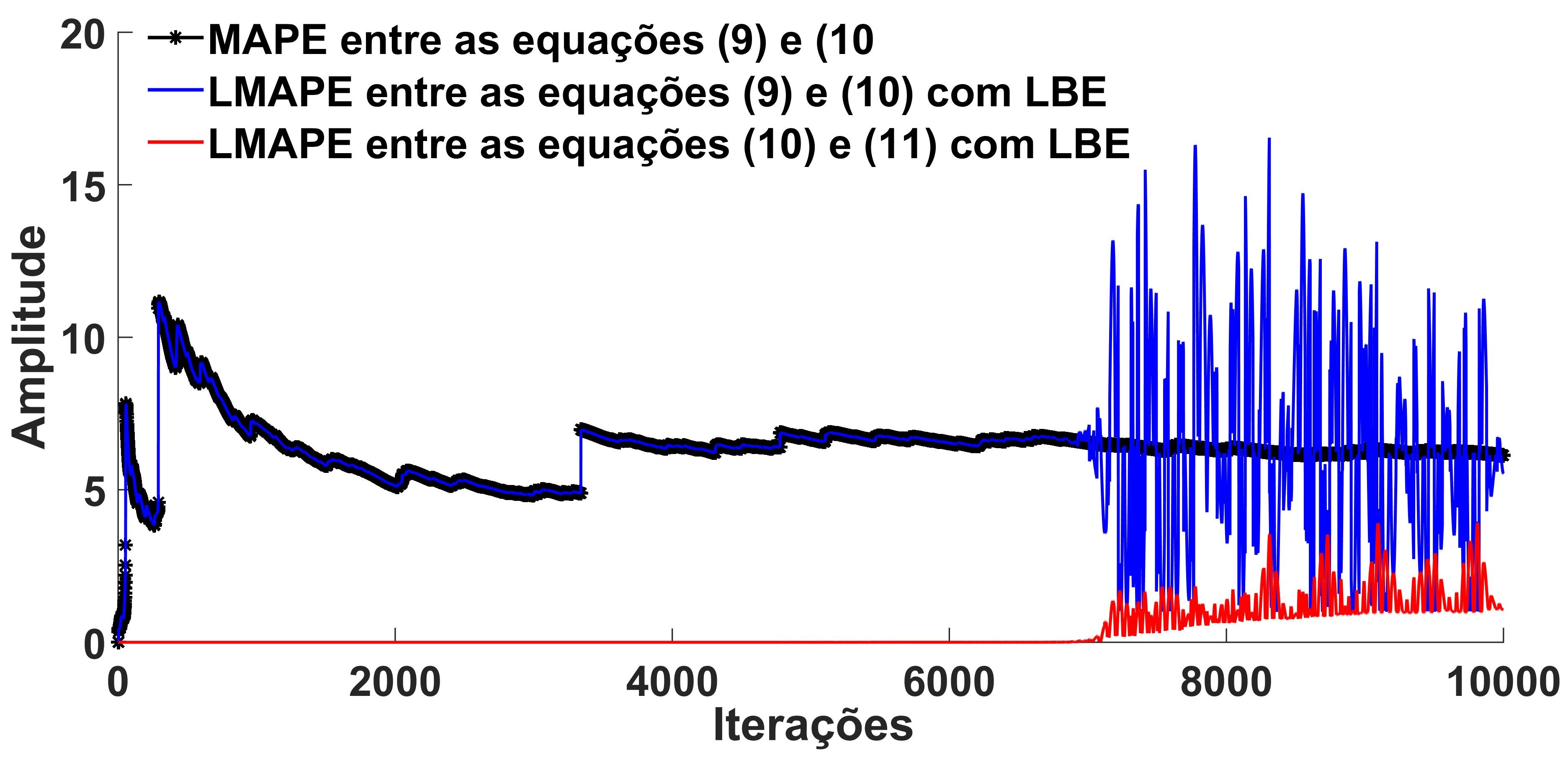}}
\caption{Análise dos resultados para os dois índices referente ao Oscilador de Duffing.}
\end{figure}

\begin{multicols}{2}
\scriptsize
\noindent\begin{eqnarray}  
  A cos(t)& = & \frac{d^2y}{dt^2}+k\frac{dy}{dt}+\mu y^3 \\ 
   G(X_n) &= &\underline{2,1579X_n-1,3203X_{n-1}}\\ \nonumber
             & & \underline{ +0,16239X_{n-2}}+0,000341U_n\\ \nonumber
             & &+0,0019463U_{n-1}-0,0048196X_n^3\\ \nonumber
             & &+0,003523X_n^2X_{n-1}\\ \nonumber
             & & -0,0012162X_nX_{n-1}X_{n-2} \\ \nonumber
             & &+0,0002248X_{n-2}^3  
\end{eqnarray}
\noindent\begin{align}
      H(X_n) =& 0,0003416U_n+0,0019463U_{n-1} \\ \nonumber
           &\underline{+2,1579X_n -1,3203X_{n-1}} \\ \nonumber
           &\underline{+0,16239X_{n-2}}-0,0048196X_n^3 \\\nonumber
           &+0,003523X_n^2X_{n-1}\\\nonumber
           &-0,0012162X_nX_{n-1}X_{n-2} \\\nonumber
           &+0,0002248X_{n-2}^2 
\end{align}

\end{multicols}

\begin{comment}
\begin{minipage}{6cm}
\footnotesize
\begin{align}
    G(X_n) = &\underline{2,1579X_n-1,3203X_{n-1}}\\ \nonumber
              & \underline{ +0,16239X_{n-2}}+0,000341U_n\\ \nonumber
              &+0,0019463U_{n-1}-0,0048196X_n^3\\ \nonumber
              &+0,003523X_n^2X_{n-1}\\ \nonumber
              & -0,0012162X_nX_{n-1}X_{n-2} \\ \nonumber
             &+0,0002248X_{n-2}^3  
\end{align}
     
     \nonumber 

\end{minipage}\hfill
\begin{minipage}{7cm}
\footnotesize
   \begin{align}
    H(X_n) =& 0,0003416U_n+0,0019463U_{n-1} \\ \nonumber
           &\underline{+2,1579X_n -1,3203X_{n-1}} \\ \nonumber
           &\underline{+0,16239X_{n-2}}-0,0048196X_n^3 \\\nonumber
           &+0,003523X_n^2X_{n-1}\\\nonumber
           &-0,0012162X_nX_{n-1}X_{n-2} \\\nonumber
           &+0,0002248X_{n-2}^2 
   \end{align}
    \nonumber 

\end{minipage}
\end{comment}

Na Figura (2), o índice RMSE tende a permanecer em uma região. O $\mathrm{LRMSE}$, quando compara o sistema com o modelo, através das Equações (9) e (10), respectivamente, tende ficar oscilando entre os valores do RMSE, tendo uma diferença de 11,2 $\times 10^{-14}$\% na 65ª iteração. Quando comparando o modelo com a extensão, por meio das Equações (10) e (11), respectivamente, o $\mathrm{LRMSE}$ tende a permanecer próximo de zero e posteriormente crescer, possuindo uma diferença de 99,99\% na 65ª iteração. Os métodos evidenciam que os valores reais localizam-se acima dos dados obtidos, além de comprovar que mesmo obtendo um modelo fiel ao sistema, ao realizar simulações computacionais, geram erros de simulação como mostrado através da comparação entre  as Equações (9) e (10). As mesmas observações podem ser realizadas para o $\mathrm{LMAPE}$, no qual apresentam uma diferença de 15,1$\times 10^{-14}$\% e 99,99\%, respectivamente para a 65ª iteração.

% Duffing 

% Sistema e modelo:  RMSEl = -3.552713678800501e-15  -0.000000000000112%  MAPEl = -1.065814103640150e-14  -0.000000000000151%
% Modelo e extensão: RMSEl =  3.173046336058120       99.999999999999943%    MAPEl =  7.040136941852619   99.999999999999787
%

\section{Conclusões}

A não consideração dos erros computacionais pode resultar em resultados relativamente incoerentes, induzindo assim, a conclusões parcialmente corretas. No entanto, com a redefinição dos métodos de validação de modelos, foi possível verificar um comportamento mais realista dos sistemas investigados, além de evidenciar que os resultados finais obtidos delimitam a região miníma em que os dados reais estão localizados. Pode-se ainda ressaltar a diferença dos resultados apresentados pela simulação do $\mathrm{LRMSE}$ e $\mathrm{LMAPE}$ comparando os sistemas e modelos e comparando os modelos e as extensões. Essa diferença se dá pois, embora representem fielmente o sistema e sejam matematicamente equivalentes, os modelos e extensões requerem operações computacionais distintas para serem simulados \cite{nepomuceno2017}. Para trabalhos futuros, pretende-se englobar aos índices de validação a questão das operações matemáticas nos computadores.

\section{Agradecimentos}
Agradecemos à CAPES, CNPq e FAPEMIG pelo apoio e à UFSJ.

\end{document}